\documentclass{PoS}

\title{QCD plasma instability and thermalisation at heavy ion collisions}

\ShortTitle{QCD plasma instability}

\author{Dietrich B\"odeker\\
  Fakult\"at f\"ur Physik, Universit\"at Bielefeld, 
  D-33615 Bielefeld, Germany\\
  E-mail: \email{bodeker@physik.uni-bielefeld.de} }

\author{\speaker{Kari Rummukainen}%
\\
  Department of Physical Sciences,
  P.O.Box 3000, 90014 University of Oulu, Finland\\
  E-mail: \email{kari.rummukainen@oulu.fi}}

\abstract{Under suitable non-equilibrium conditions QCD plasma can
  develop plasma instabilities, i.e. exponential growth of some
  modes of the plasma.  It has been argued that these
  instabilities can play a significant role in the
  thermalisation of the plasma in heavy-ion collision experiments.  We
  study the instability in SU(2) plasmas using the hard thermal loop
  effective lattice theory, which is suitable for studying
  real-time evolution of long wavelength modes in the plasma.  We
  observe that under suitable conditions the plasma can indeed 
  develop an instability which can grow to a very large magnitude,
  necessary for the rapid thermalisation in heavy-ion collisions.}

\FullConference{The XXV International Symposium on Lattice Field Theory\\
		 July 30-4 August 2007\\
		 Regensburg, Germany}

\usepackage{bm}
\newcommand{\lsi}{\raise0.3ex\hbox{$<$\kern-0.75em\raise-1.1ex\hbox{$\sim$}}}
\newcommand{\gsi}{\raise0.3ex\hbox{$>$\kern-0.75em\raise-1.1ex\hbox{$\sim$}}}
\newcommand{\lsim}{\mathop{\lsi}}

\renewcommand{\vec}[1]{{\bm #1}}

\newcommand{\be}{\begin{equation}}
\newcommand{\ee}{\end{equation}}

\begin{document}

\section{Introduction}

\begin{figure}[t]


\centerline{\includegraphics[width=0.5\textwidth]{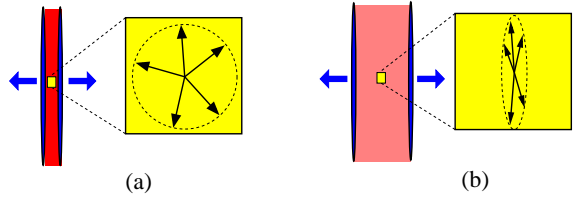}}

\caption{The longitudinal expansion of the collision
  volume makes the initial parton momentum distribution (a)
  squeezed along the plane perpendicular to the collision axis (b).}
\label{fig:asym}
\end{figure}

One of the most striking results from the heavy ion collision
experiments at RHIC is the rapid thermalisation of the plasma; the
thermalisation appears to occur in time $ \lsim 1\mbox{fm}/c$ after
the collision \cite{early}.  At sufficiently large collision energy
the QCD coupling is small and perturbation theory should be
applicable.  However, it turns out that the perturbative processes
cannot alone explain rapid thermalisation \cite{bottomUp,impact}.  
It has been argued that the strongly non-equilibrium initial
conditions may lead to exponential growth of certain long wavelength
modes of the plasma --- {\em plasma instability} \cite{Mrowczynski}.
These growing modes might play a significant role in the
thermalisation of the plasma.  The plasma instability arises because
the plasma initially expands predominantly along the collision axis
($\hat z$ direction), and the momentum distribution of the produced
partons becomes anisotropic: the momentum distribution becomes much
smaller along $z$-axis direction than along the transverse directions,
no matter what the initial distribution of the partons was
(Fig.~\ref{fig:asym}).  The initial momenta of the partons is of
order of $\sim$ few GeV (which is the saturation scale of the 
original nuclei in color glass condensate models), which we 
denote as the ``hard'' scale.  

\begin{figure}[t]


\centerline{\includegraphics[width=0.33\textwidth]{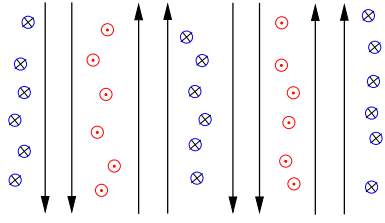}}

\caption{The Weibel instability in the electromagnetic plasma.  Arrows
  show the electric current, circles the magnetic flux perpendicular
  to the plane.  The magnetic field amplifies the inhomogeneities in
  the current, which in turn amplifies the magnetic fields.}
\label{fig:current}
\end{figure}

The hard partons will interact with the soft gauge fields; 
assuming that the soft
fields have small initial amplitude 
the non-abelian nature of the fields can be ignored. 
In this case the anisotropic parton
momentum distribution causes the soft fields to become 
unstable against the generation of $\hat x$
and $\hat y$ -direction magnetic fields: the magnetic fields focus the
current carried by the partons by amplifying the inhomgeneities in it,
which in turn leads to increasing magnetic fields.  This leads to
exponential increase in the magnitude of the magnetic fields,
analogously to the Weibel instability in electromagnetic plasmas
(Fig.~\ref{fig:current}).  However, for the case of QCD the current is
mostly carried by saturation scale partons, which are mostly 
hard gluons.

The growth in the small-field regime happens only in a certain range
of wave vectors (depending on the degree of anisotropy in the hard
parton distribution) and it is maximal at a particular wave vector,
$\vec k_*$, oriented along $\hat z$-direction.  In QED the growth can
continue until the magnitude of the gauge field reaches $e A \sim
p_{\rm hard}$; when this happens the hard charged particles are
deflected to random directions and their distribution becomes
isotropic.  However, in QCD the field equations become non-linear at
much smaller magnitude $g A_{k_*} \sim k_*$ (or $B^2 \sim g^2 k_*^4$),
because $k_* \ll p_{\rm hard}$.  Thus, the central question is what
happens to the unstable growth when the magnitude of the
chromomagnetic fields reaches this ``non-abelian'' value.  In
Ref.~\cite{arnoldAbelianization} it was suggested that the growth
could persist beyond the non-abelian value if the system
``abelianises,'' i.e. it becomes essentially dominated by only one
color degree of freedom.  Thus, as the fields continue growing the
distribution can isotropize through the mechanism described above.

Because of the large amplitude of the chrmomagnetic fields the
problem is non-linear and non-perturbative.  The cleanest way to
approach the problem is to perform real-time evolution on the lattice
using so-called ``hard loop approximation'': the infrared modes are
classical chromomagnetic fields, and the hard partons are treated as a
classical charged particle current on the soft field background.  This
approximation is justified because we will be dealing with large
occupation numbers for the soft fields, and the expansion renders the
hard particle distribution dilute.  We also consider only
non-expanding systems with fixed anisotropic hard particle momentum
distributions in order to focus on the effects of the anisotropy.
Physically this corresponds to sufficiently large times where the
expansion rate is parametrically small compared to the rates
associated with the instability.

This approach has previously been applied to 1+1 -dimensional case
\cite{romatschke2d} where it was observed that the fields indeed
continue to grow in the non-linear regime.  However, 3+1 dimensional
simulations with moderate anisotropies have indicated that the
instabilities are quenched as the non-linearities become important
\cite{arnoldFate,romatschkeFate}.  In this work we shall study
considerably stronger momentum anisotropies than above, together with
large lattice volumes and small lattice spacing.  A detailed
report of the results can be found in \cite{dbkr}.  Strong
anisotropies are also studied in Ref.~\cite{arnoldExtreme}, but
with initial conditions not leading to further exponential growth.

\section{Hard Loop effective theory}

The hard modes are described as on-shell particles moving in 
soft background fields, with a distribution function 
\be
  f_{{\rm hard}}(x,\vec p) = \bar f( \vec p ) + 
  \lambda^a f^a(x, \vec p ) + \ldots
\ee
where the anisotropic gauge singlet part $\bar f(\vec p)$ 
we assume to be constant in space and time, and 
$f^a$ describes fluctuations in the current carried by the particles.
The system evolves according to the Yang-Mills-Vlasov equations
of motion
\begin{equation}
  ( D _  \mu    F ^  {\mu  \nu  } ) ^ a = J^{a,\nu}_{{\rm hard}}
  = g \int_{\vec{p}}\, v ^ \nu f ^ a,
~~~~~~~~~~~~
  ( v \cdot D f ) ^ a + g v ^ \mu  F _{\mu  i } ^ a 
  \frac{ \partial \bar{f} }{\partial    p ^ i} = 0\,,
  \label{vlasov}
\end{equation}
where $v = (1,\vec p/p)$.  Defining 
\be
  W ^ a ( x, \vec v ) \equiv 4 \pi  g \int\limits _ 0 ^ {\infty  }
  \frac{  d  p  p ^ 2}{(2 \pi )^ 3} f ^ a ( x, \vec p )
\ee
we can integrate the equations of motion over $|p|$, obtaining 
\begin{equation}
  ( D _  \mu    F ^ {\mu  \nu  } ) ^ a =
  \int \frac{ d \Omega_{\vec v}}{4 \pi }
  v ^ \nu  W ^ a,
 ~~~~~~~~~~~~~~~~~~
  ( v \cdot D W ) ^ a =   m_0^2 \, 
       v ^ \mu  F _{\mu  i } ^ a  U^i(\vec v). \label{eqm}
\end{equation}
Here the vector $U^i(\vec v)$ characterises the anisotropic singlet part 
of the hard distribution $\bar f$:
\be
  m_0^2 U ^ i (\vec v ) = -4 \pi  g ^ 2  \int\limits _ 0 ^ {\infty  }
  \frac{  d  p  p ^ 2}{(2 \pi ) ^{3}}  \frac{ \partial \bar{f } ( p \vec v ) }{\partial p ^ i}.
\ee
For thermal distribution $\bar f$ becomes isotropic, and we would
obtain $\vec U = \vec v$ and $m_0 = m_{\rm Debye}$, the Debye mass of
the thermal plasma.  We note that $m_0$ is the only dimensionful
parameter in the problem.

\begin{figure}[t]


\centerline{\includegraphics[width=0.4\textwidth]{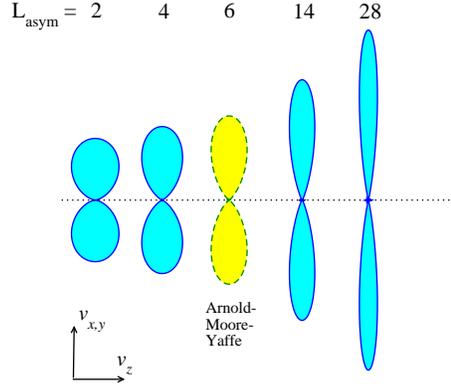}}

\caption{Anisotropic hard particle distributions used in
  this work, together with the distribution used 
  by Arnold, Moore and Yaffe~\cite{arnoldFate}.  The
  distributions are plotted so that the relative number of
  particles moving to direction $\vec v$ is proportional
  to the length of the radial vector from the center of the
  plot.}
\label{fig:asym_polar}
\end{figure}

The equations of motion \ref{eqm} are discretised on the lattice.
The current carried by the hard particles is described by the
$W^a(x,\vec v)$-fields.  These are quite expensive to handle, because
they live on manifold $R^3\times S^2$.  We treat these by expanding
the distributions in spherical harmonics:
\begin{equation}
   W^a(x,\vec v) =
     \sum_{\ell m} W^a_{\ell m} Y_{\ell m}(\vec v),
 ~~~~~~~~~~~~~~~~~~
 \bar f(\vec p) = \sum_\ell \bar f_{\ell}(p)Y_{\ell\,0}(\vec v),
\end{equation}
where $\ell = 0\ldots L_{\rm max}$, the cut-off in spherical harmonics
expansion.  Thus, at each site $W^a$ has $(L_{\rm max} +1)^2$ real
degrees of freedom.  This approach has been also used in
Refs.~\cite{arnoldFate,arnoldExtreme} to study the plasma
instablity.  Originally, this method was successfully applied to the
calculation of the sphaleron rate in hot SU(2) gauge theory on 
the lattice \cite{lmax}.

For simplicity, we are using SU(2) gauge group in our analysis.  We
present the results using 4 different values for the anisotropy of the
$\bar f$, both weaker and much stronger than used in
\cite{arnoldFate}.  Each distribution is characterised by the maximal
spherical harmonic index used to parametrise $\bar f$, $L_{\rm asym}
=$ 2, 4, 14 and 28 ($L_{\rm asym} < L_{\rm max}$).  For each value of
$L_{\rm asym}$ we approximately maximised the possible asymmetry of
the distribution; the motivation for this is that this choice should
minimise the required $L_{\rm max}$ cutoff.  The anisotropic
distributions are shown in Fig.~\ref{fig:asym_polar}.  The degree of
the anisotropy is characterised by the {\em anisotropy
  parameter} $\eta^2 \equiv 3\langle v_z^2 \rangle/\langle v^2
\rangle$; for the distributions here this is $\eta^2 =$ 0.6, 0.4,
0.086 and 0.022 for $L_{\rm asym}=$ 2, 4, 14 and 28.

In our simulations we are using very large lattice volumes 
(up to $240^3$) and vary the lattice spacing by more than order
of magnitude.  The $L_{\rm max}$-cutoff is up to 48.  In general, the
infinite volume and continuum limits are under control; for details,
see \cite{dbkr}.

\begin{figure}[t]
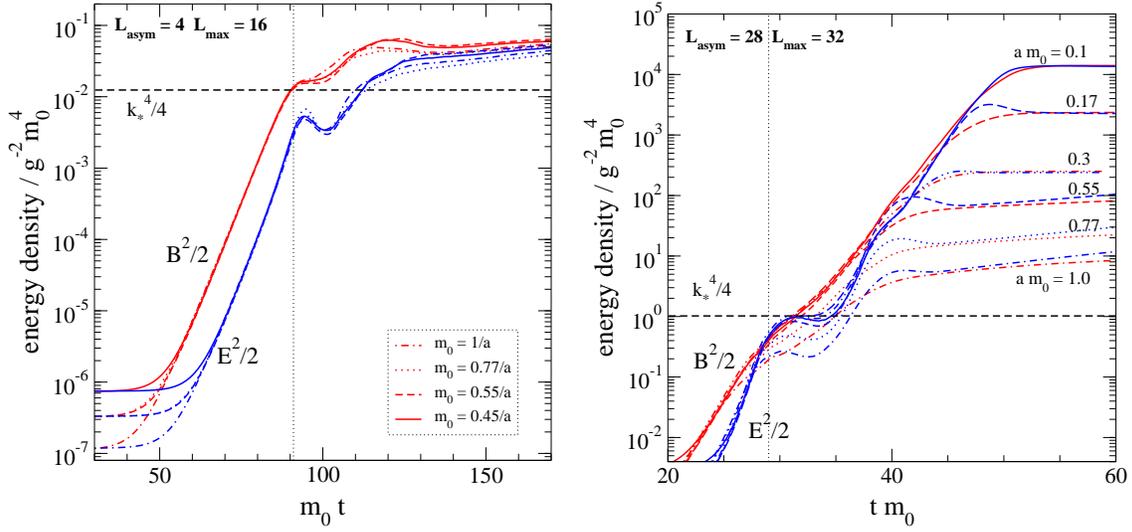



\centerline{
\includegraphics[width=0.48\textwidth]{w3d_L4l16_mcmp.eps}
~~
\includegraphics[width=0.48\textwidth]{w3d_L28l32_mcmp.eps}
}

\caption{Growth of the energy with small (left) anisotropy, $L_{\rm
    asym}=4$
and large (right) anisotropy, $L_{\rm asym} = 28$.
In both cases the energy grows until the magnitude of the
growing magnetic field reaches the value where non-abelian
effects become significant; $B^2 \sim (k^*)^2$.  
With strong anisotropy, the growth continues until regulated by
the lattice cutoff.}
\label{fig:growth}
\end{figure}

\section{Results}

In Fig.~\ref{fig:growth} we show the growth of the soft field energy
density at small and large anisotropy, measured at different lattice
spacings.  Initially the soft fields have small white noise
fluctuations.  In both cases the instability causes exponential growth
of energy density in the linear (weak field) region.  However, when
the field evolution becomes non-linear (shown as vertical lines), the
growth is rapidly quenched at weak anisotropy, independent of the
lattice spacing.  This is in accord with the results of
Ref.~\cite{arnoldFate}.

\begin{figure}[t]


\centerline{
\includegraphics[width=0.48\textwidth]{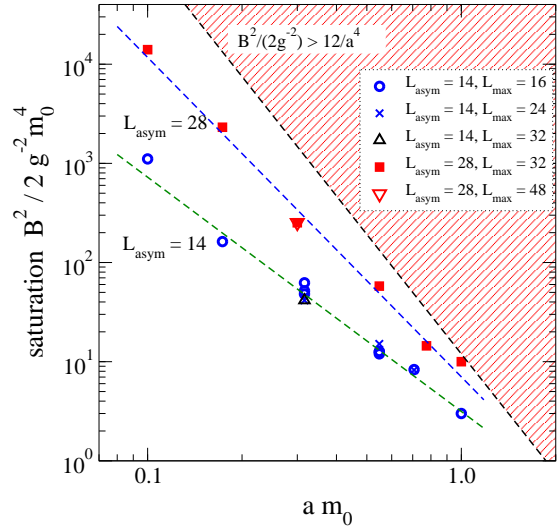}
}

\caption{Saturation magnetic field energy as
a function of the lattice spacing at large anisotropies.  The
red hashed area is forbidden because energy density there is too large to
be supported by the lattice.}
\label{fig:saturation}
\end{figure}

However, at strong anisotropy the growth continues in the non-linear
regime, and the smaller the lattice spacing is, the further the growth
persists.  The cutoff is due to lattice cutoff, as can be seen in
Fig.~\ref{fig:saturation}: here we show the chromomagnetic field
energy density at final saturation as a function of the lattice
spacing.  The saturation energy is well described as a power law of
the lattice spacing.

\begin{figure}[t]


\centerline{
\includegraphics[width=0.48\textwidth]{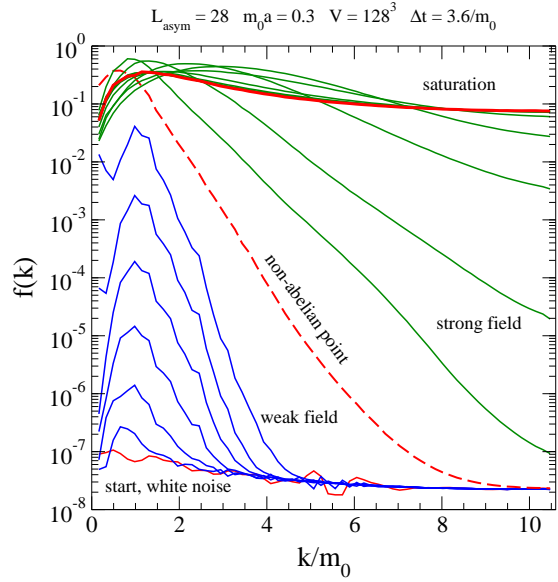}
}

\caption{
Coulomb gauge power spectrum (occupation number)
as a function of time for strong anisotropy.  The spectra
are plotted at time intervals of
$\Delta t = 3.6/m_0$, with the initial state at bottom, and the final
state near the top.
}
\label{fig:spect}
\end{figure}

Thus, the results clearly indicate that unstable growth is possible in
the non-linear regime.  What field modes do grow here?  We study this
by fixing to Coulomb gauge and measuring the occupation numbers of the
gauge field, $f(\vec k) \propto |\vec k| A(\vec k)$.  The evolution of
the occupation numbers at large anisotropy is shown in
Fig.~\ref{fig:spect}.  In the linear (early) regime the growth near
$k_* \approx m_0$ is clearly visible.  However, when the system
becomes non-linear at $f(k_*) \sim 1$, the growth at $k_*$ is
completely quenched, but $f(k)$ at higher wave numbers shoots rapidly
up.  The final occupation number distribution is very close to the
thermal one.  Thus, the growth mechanism appears to be very different
from the abelianisation proposed in \cite{arnoldAbelianization}.  We
have checked this behaviour using various gauge invariant measurements
($k$-sensitive operators, cooling), with fully consistent results, see
Ref.~\cite{dbkr}.  Unstable growth in the non-linear regime has also
been observed in Ref.~\cite{dumitruAvalanche}, but using very
different methodology.

In summary, we observe clear signal of rapid soft field energy growth
in the non-linear (large magnitude) regime when the hard particle
distribution is strongly anisotropic, suggesting possible role in the
thermalisation of the plasma in heavy ion collision experiments.
However, the mechanism through which the growth proceeds is still
unknown and under further study.  There are also important caveats:
perhaps most significantly, the initial conditions in the cases
reported here all have small magnitude soft fields.  When the
magnitude of the initial fields is increased the non-linear growth is
reduced \cite{dbkr,arnoldExtreme}.

DB acknowledges support from DFG funded Graduate School GRK 881, and
KR support from Academy of Finland grants 104382 and 114371.  The
simulations in this work have been performed at the Finnish IT Center
for Science (CSC, Espoo, Finland).

\end{document}